# Digital Library Initiatives in North East India: A Survey

**Sankhayan Mukherjee & Swapan Kumar Patra**[1]

**Department of Library & Information Science**

**Sidho-Kanho-Birsha University**

**Purulia, 723149, West Bengal, India**

**Email: skpatra@gmail.com; skpatra@skbu.ac.in**

## Abstract

This is a survey of digital library initiative of North East India. The recent initiative by the government of India towards the digitization is reflected in various digitation programs. The secondary sources of data are used to map the 16 digital library initiatives in eight north east state of India. The study has observed that digital library in true sense is perhaps lacking. Many of the digital libraries are not accessible from the outside and lack regular maintenance. In this context, a national level policy initiative is the need of the hour including various stakeholders from the academics, library professionals etc. The study also comes up with various important observations and policy suggestions which may be helpful for scholar, librarians, policy and decision makers in the government.

[1] Corresponding Author

## Introduction

A special library that mainly focuses on collection, storage and retrieval of digital objects like text, video material, photograph, audio material etc. in electronic media formats (as opposed to print, microform etc.), may be called as a Digital Library (Balasubramanian 2021). The size and facility of a Digital Library may vary depend on the need and circumstances. It can be maintained by individuals, associations, a physical library or an academic institution. The electronic content can be accessed locally or remotely through the Internet.

The concept of digitalization of library materials or the concept of digital library in India was started in mid 1990s. With the development of information and communication technology (ICT) and particularly the Internet technology in Indian market, the concept of digitalization was also become popular. In 1996, at a conference of Society of Information Science, Bangalore, the concept of digital library was discussed (Gurram 2008). However, the concept did not get that much of importance among the library organizations at that time. A few years later that the concept was many takers. However, majority institutions or organizations 'digital library' is mainly staying with subscribing e-journals, scanning document and upload them on the Intranet. This model is perhaps is far away from the actual definition of 'digital library' which may include a proper digitized collection with a strong metadata creation and information retrieval mechanism from the digital collection.

The digital library initiative is getting popular in India because of the information technology and easy Internet access. Moreover, Government of India has taken several initiatives to digitize various resources. As a result, Digital Library of India (DLI) is the biggest national level initiative in India taken by the Indian government. It is a part of the Universal Digital Library Project, regarded by Carnegie Mellon University, USA. Other than that Ministry of Education under its National Mission on Education through Information and Communication Technology has initiated the National Digital Library (NDL) project to develop a framework of virtual repository of learning resources with single-window search facility. It is being developed at IIT Kharagpur. Beside this, *Digital Library of art Masterpiece, National Mission for Manuscripts*, *Vidyanidhi*, *National Institute of Advanced Studies* (NIAS), Bangalore, etc are also noteworthy digital library initiatives in India.

Along with the above mentioned major digital library indicatives' there are many digital library started by academic, government and other agencies all over India. This study is an investigation of digital library initiatives of North-East India. While doing so, this study will investigate the following research questions: What is status of digital library initiatives of eight states (Assam, Arunachal Pradesh, Mizoram, Manipur, Meghalaya, Nagaland, Sikkim and Tripura). These North-East states of India are generally regarded as comparatively less developed regions of India. Hence this investigation will perhaps be useful for the researchers in library or information science as well as policy and decision makers.

## Literature Review

Jeevan (2004) highlighted that digitization can be taken as a visible proposition to extend the shelf life of non-digital content through preservation. Hence, the uses of digital library will be increased as well. Looking at the importance of digitization there are many initiatives taken all over the world to digitize various types of contents. The developing countries around the world have taken initiatives to digitize the contents. Kaur, Singh (2005) observed that, in developed countries 60%-70% data is available in digital format. Whereas, in developing country like India there is only 2.5% of data available in digitized format. So, they discussed about the need for a National Information Policy and the training of the library professionals to give the velocity of transformation of traditional library into digital library.

However, Jeevan (2004) has observed that for majority of Indian libraries, the 'digital library' is a concept limited to the computerization of internal operations (using databases on electronic media such as CD-ROMs), web access to subscribed journals and various free resources. However, what is lacking, especially in developing countries, is an integrated collaborative approach to identify valuable content for digitization and bring institutions with adequate financial and infrastructural support. Digital library development in India requires two-pronged strategy including both the digitization of local content in one hand and access to external resources on the other hand.

Problems in digital library development are manifold in India such as lack of interest, non-availability of computers and IT infrastructure for library activities, copyright issues, ensuring safe access, properly selecting materials from the vast amount available, internet bandwidth, and absence of adequate funds. Sreekumar & Sreejaya (2005) further emphasized the issues in the initiative of digital libraries in India. They mainly focused on the problems, that are, proper infrastructure of ICT and their policies, proper planning and integration of information resources, lack of flexibility of publisher's policies and data format, lack of management and technical skills & the IPR issue.

It is a very promising thing that, now a days, Government of India and various state governments focus on the development and activities of the digital libraries in India. Bhattacharya (2004) explained the earlier initiatives, such as the INDEST Consortia and so on including the challenges faced by a digital library, problems during development of a digital library and the digital divide that faced by the country. Government of India has taken several initiatives to establish digital library in the country. The Ministry of Human Resource Development (MHRD) took various initiatives such as "Consortia-based Subscription to Electronic Resources for Technical Education System" and named it as Indian National Digital Library in Science and Technology (INDEST) Consortium (Arora & Agrawal 2003) Other than the government, The Energy and Resources Institute (TERI) Digital library provides scholars to better single window access to structured information from their desktop (Deb 2006).

A number of research papers deal with the various institutional level digital library initiatives. These types of studies may be considered as case study of digitization. For example, Gaur (2003) dealt with the digitalization status in Indian Management Institutes in India which mainly includes the 'library automation'. Building procedure of digital resources in the Indian National Science Academy (INSA), New Delhi is discussed in a paper by Mujoo-Munshi (2003). Joshi (2006) has discussed about the Tocklai Central Library digitaization initiative of the institute. Naga, Pradhan, Arora & Chand (2008) studied about the usage trends of e-journals in universities in North-East in India and observed that there is a desperate need to improve the service of the universities to provide a good resource(e-journals) to meet the users' need.

As observed form the literature review above, there are many studies available of digital library initiatives in India. All these studies are mainly case study of establishment of digital library in a single institute. A comprehensive study of digital library covering the Indian scenario as a whole is lacking. Moreover, the literature review has not found any significant study on the digital library initiative of north eastern part of India. Also, there is no study which is dealing with the issues, for example, copyright issues, metadata creation, retrieval efficiency and so on. Hence this study will fill the gap in the existing status of digital library in India from the North East states of India.

## Objectives of the study

The Indian government launch the digital library initiative to foster creativity of citizens and to get the free access to the all-human knowledge. So, with this idea in mind the goal of the study is a survey of various digital library initiatives of North Eastern States of India from their respective websites. This study is going to survey the digital library initiatives by investigating the following issues: The number of collections, the access mode, software used for digitation, users' statistics, copyright issues, related collection development and usage policies and third-party access etc. The above mention parameters will be useful to get an insight of the current trends of digital library initiative of North East India.

## Methodology

This study is based on the secondary sources, particularly the respective digital library initiative's web pages. The websites were accessed in the month of July to September 2022. There are altogether 16 various digital libraries investigated in this study. The libraries are considered for this study are from the following North Eastern states: Arunachal Pradesh (3) Assam (6), Meghalaya (1), Nagaland (1), Sikkim (1), Manipur (1), Mizoram (1), Tripura (1). The details sample size is shown in Table 1. An exhaustive literature search was carried out and the respective libraries' website was visited and the findings are present in the following sections.

## Results

From the secondary literature, it was observed that articles in this area have highlighted on the digitization process and the digitization effort of the institutions. Although some of those were touched the policy and copyright issues during the digitization. However, there was no reasonable study about 'Digital Library initiative in North-East India'. We get the following results during the study of digital library initiative in North-East India. The state wise distribution of the digital library initiative is presented in Table 1.

**Table 1 Various digital Library initiatives of North East India.**

| State | Digital Libraries | Website Link for corresponding Institutions |
|---|---|---|
| Arunachal Pradesh | 1. Rajiv Gandhi University Library | https://rgu.ac.in/library/ |
| | 2. Virtual Library Solution, Government of Arunachal Pradesh | http://referencelibraryresearch.com/ |
| | 3. Arunachal University of studies | https://www.arunachaluniversity.ac.in/pages/library |
| Assam | 1. Digital Library of Assam<br>Directorate of Library Services<br>Department of Cultural Affairs<br>Government of Assam | https://publiclibraryservices.assam.gov.in/ |
| | 2. Digital Library: Raj Bhavan, Assam | http://rbassam.digitallibrary.co.in/ |
| | 3. Assam District Library, Guwahati | https://publiclibraryservices.assam.gov.in/ |
| | 4. National Law University and Judicial Academy Digital Repository cum Digital Library | http://librarynlujaa.blogspot.com |
| | 5. Diphu Government College | http://diphugovernmentcollege.com/DGC/CentralLibrary.action |
| | 6. Central Institute of Technology, Kokrajhar | http://centrallibrary.cit.ac.in/ |
| Meghalaya | Central Library, North-Eastern Hill University | https://www.nehu.ac.in/lib rary/index.html |
| Nagaland | Central Library: Nagaland University | https://library.nagalanduniversity.ac.in/?q=Digital%20Libraries |
| Sikkim | Teesta-Indus Central Library, Sikkim University | https://library.cus.ac.in/index.php/subject-wise-e-books/ |
| Manipur | Manipur University Library | https://mulibrary.manipuruniv.ac.in/ |
| | Manipur Technical University Library | http://mtu.ac.in/library/ |
| Mizoram | Mizoram Sate Library | https://statelibrary.mizoram.gov.in/page/digital-library |
| Tripura | Birchandra State Central Library | https://bcscl.tripura.gov.in/ |

Although Table 1 is perhaps is not a comprehensive list of digital libraries in North East India but it is the list of the prominent digital library of NE India. This study has considered academic, government and private sector's institutions where there is a mention of the digital library on the website.

**Table 2 An Assessment of various digital libraries in North East India using various parameters**

| Sl No. | Digital Libraries | No. of Collections | Access Mode | Software | Users' Statistics | Copyright Issues | Policies (if any) | Third Party Access (If any) | Comments |
|---|---|---|---|---|---|---|---|---|---|
| 1 | Rajiv Gandhi University Library<br>Arunachal Pradesh | NA | Closed Access | KOHA | NA | | NA | Ebsco Discovery Service and NDLI | The library has social media presence. It has its Facebook page: https://www.facebook.com/groups/rgu.library |
| 2 | Virtual Library Solution<br>Arunachal Pradesh | More than 10,610 | Open Access | NA | Visitor : 4948 | | NA | NA | The library has Web OPAC |
| 3 | Arunachal University of studies<br>Arunachal Pradesh | NA | NA | NA | NA | | | Access Through NDLI | The website has a specific section for its own digital library, it is very surprising that there is no link to open digital library from that section |
| 4 | Digital Library of Assam<br>Assam | Total Books: 3,453 | Close Access | NA | NA | Copyright free books | **Copyright Policy** Material featured on this Website may be reproduced free of charge after taking proper permission | NA | Common people can't access this site. Actually, there is no user guide for accessing the e-resources, although there is an e-book list in the website |
| 5 | Digital Library: Raj Bhavan, Assam | 239, till 15th Sept ,2022 | Open Access | DSpace | Visitor:123806 | Raj Bhavan Library, Assam © 2021. All rights reserved. | NA | NA | The Library have its Web OPAC<br>Mainly we can access the Govt. of Assam notifications, press releases etc form this site |
| 6 | District Library, Guwahati | | | | | | | | There is a particular section for digital library in the site, but it leads to the Digital Library of Assam's website. |
| 7 | NLUJAA Digital Repository cum Digital Library | NA | Close Access | KOHA | Website Page views 20,535 till 15th Sept,2022 | | The NLUJAA Library's mission is to support legal teaching learning and research of the university community by providing comprehensive resources and services in the subject of Law and allied field. | Enhanced by Google | Has it's own Web OPAC named NLUJAA OPAC |
| 8 | Diphu Government College | NA | Close Access | NA | NA | | NA | NA | The library has its Web OPAC |
| 9 | Central Institute of Technology, Kokrajhar | 14,260 | Close Access | Soul 2.0 | NA | | NA | Knimbus | The library has its Web OPAC |
| 10 | NEHU Central Library | NA | Open Access | NA | NA | | NA | Knimbus, NDLI | The library has its Web OPAC |
| 11 | Central Library: Nagaland University | NA | Close Access | NA | NA | | **Copyright Policy** Material featured on this website may be reproduced free of charge after taking proper permission | NA | There is remote access facility in the library |
| 12 | Teesta-Indus Central Library, Sikkim University | NA | NA | DSpace: Digital Repository System KOHA: Integrated Library System IRINS – Faculty Research Profile Remote Access (IDP) | NA | NA | NA | NA | Have its' own Central Library app |
| 14 | Manipur University Library | NA | Close access | NA | 41719 vistors | Electronic resources such as e-journals, e-databases, e-books made available by the MU Library are for academic use | Copyright Policy: The use of robots, spiders, or intelligent agents to access, search and/or systematically download from these resources is prohibited. Any violation of this policy will result in penal action as per the rules and regulations of the university. | NDLI, NPTEL etc | The Library has its own Web OPAC |
| 15 | Manipur Technical University Library | NA | Open Access | The Library uses Koha software which is an integrated multi-user library automation management system that supports all in-house activities of the library. | NA | NA | The main objective of the Library is to support the research and educational activities of the Institute. | World eBook Library, NDLI etc. | Not a digital library in true sense, some digital collections can be accessed through this website. |
| 16 | Mizoram Sate Library | A total of 316 rare and copyright books are digitized which is available for access in the library. | Close Access | LIBSYS | Total Visitors 887 till 29th Sep 17, 2:52 PM | NA | NA | NA | The library has its Web OPAC |
| 17 | Birchandra State Central Library | NA | Open Access | NA | NA | NA | **Copyright Policy** Material featured on this Website may be reproduced free of charge after taking proper permission | Project Gutenberg, Digital Library of India) Google Books Open Library Free E-Books.net | It does not own a digital library, but some e-resources can be accessed through some third-party site from this website. |

This study considers various parameters to evaluate the digital libraries of North East India. These parameters are; number of collections of the library, access mode of the library, archiving software that used by the library, users' statistics, copyright issues, policies, and third-party access and so on. Table 2 shows that few libraries have open access (public can access it) and some have the close access system (IP base authentication or remote access user ID and password) to acquire the digital library collections. These digital repositories are using various open source as well as closed library management software like Koha, Libsys, Soul and DSpace as repository software. The institutions do not have any policy regarding digital library collection, meta data harvesting or retrieval issues. Though some libraries have their own organized policy regarding copyright policy, privacy policy etc. On the other hand, some have used third party organization such as Knimbus, Google etc. to search and retrieve their digital library collections. Moreover, a very few libraries have uploaded their users' statistics on the website. So, it is difficult to know the usages statistics from the website. The detail description of the libraries considered in this study are as follows:

**Rajiv Gandhi University Library**

The Rajiv Gandhi University (earlier it was known as Arunachal University) is the foremost institution for higher education and learning in the state of Arunachal Pradesh. The library has a collection of about 64,400 books and has subscription of various national and international journals. Moreover, the library has collection of 343 Ph.D. thesis and 353 dissertations. However, the digital collection of the library on the website is not mentioned. The library is mainly created to meet the needs of the students of the Rajiv Gandhi University. Outsiders cannot create an account in the library server for the access. The library is fully computerized and there is library automation software RFID with KOHA. The copyright issue and collection development policy are not clearly stated. The library used 'Ebsco Discovery Service' to provide service to the users. The library has a social media page and it maintains its Facebook page, where they post the latest news about their library.

**Virtual Library Solution, Government of Arunachal Pradesh**

Virtual Library Solutions is the Reference Library of the Department of Cultural Affairs, Directorate of Research, Government of Arunachal Pradesh. The library was established in the year 1976 to meet the need of researchers for the government of Arunachal Pradesh. The library has a collection of more than 10,610 books, Journals, Reports etc. The Reference Library has been completed the library automation and digitization of rare books and journals for the users and developed the library into the virtual library.

The details of library automation software, copyright issues and policies are not clearly stated. Its collection is open access for all. The website has visitors' information and has its own Web OPAC. The library website and digital collections are maintained by the third-party service providers.

**Arunachal University of studies**

We did not find anything in their website during our study. There was section for a digital library in the particular institution's website, but when we clicked on it, it did not forward it anywhere. Actually there was no link for digital library in the website. Though users can access NDLI from the website.

**Digital Library of Assam**

The library has the collection of 3,453 total digital books. Common people can't access this site. Actually, there is no user guide for accessing the e-resources, although there is an e-book list in the website. We did not find anything about the library automation software during our study. There are all copyright free books in the library. And main thing is that the library has it's own copyright policy.

**Digital Library: Raj Bhavan**

The library has 239 of total digital collections till 15$^{th}$ sept, 2022. The users can access the Government of Assam notifications, press releases etc form this library. The library uses DSpace as their repository software. And the site is open accessible for all. The website claims that they have total of 123,806 Visitors in their site. The library has their own Web OPAC.

**District Library, Guwahati**

There is a particular section for digital library in the site, but it leads to the Digital Library of Assam's website.

**National Law University and Judicial Academy (NLUJAA) Digital Repository cum Digital Library**

The library has close access system. It is mainly used by the NLUJAA students. The website displays usages statistics. The library has Koha as library automation software and has own Web OPAC named as NLUJAA OPAC.

**Diphu Government College**

This institution's digital library is called as First Digital Library initiative of North East India. But the unexpected thing is that we could not able to access the site due to the site was totally closed accessed for public.

**Central Institute of Technology, Kokrajhar**

The library has total number of 14,260 as their digital collections. It uses Soul 2.0 as their automation software. We did not find anything about users' statistics, copyright issues and policies of the digital library from the website during our study. The library has its own Web OPAC. It provides service to its users through Knimbus as third-party management.

**North-Eastern Hill University (NEHU) Central Library**

NEHU Central library does not have much information about the library collection, automation software, users' statistics, copyright issues and policies of the digital library from the website during this study. The library is open access to all and it provides service to its users through Knimbus as third-party management. The library has its own Web OPAC.

**Central Library: Nagaland University**

The library does not have information about the library collection, automation software, users' statistics, copyright issues and policies of the digital library. The library is open access and has has the remote access facility.

**Teesta-Indus Central Library, Sikkim University**

Teesta-Indus Central Library, Sikkim University uses DSpace as library repository software, Koha as library automation software. The library uses IRINS is web-based Research Information Management (RIM) service developed by the Information and Library Network (INFLIBNET) to connect to the scholarly network. The library has its own central library mobile app to provide services to its users.

**Manipur University Library**

The website does not have much information about the library collection and the library automation software. It is a close access system and has copyright policy. The library has its own Web OPAC and remote access facility is available for its digital collections.

**Manipur Technical University Library**

The library website does not have much information about the library collection, users' statistics, and copyright issues etc. The library has open access system. The library uses Koha software which is an integrated multi-user library automation management system that supports all in-house activities of the library. The website has mentioned the other e-resources and have the links of NDLI, University Press Scholarship Online (UPSO), elib4u, World eBook Library and so on through this website.

**Mizoram Sate Library**

Mizoram Sate Library is of the public library service under the Department of Art & Culture. The library has collection of a total of 316 rare and copyright books that are digitized. These digital collections are available for access in the library. The library has close access system. It uses Libsys as its automation software and its own Web OPAC. There is no information on the copyright issue and policies on the digital library in the website during our study.

**Birchandra State Central Library**

Birchandra State Central Library is located in Agartala, Tripura. It is the apex Library of the Public Library system in Tripura. There is no information the library collection, automation software, users' statistics, and copyright issues of the digital library from the website during our study. The library is open accessible for public and it has it's own policies for maintaining the library. The library has given the link to the famous digital libraries for example, Project Gutenberg, Digital Library of India, Google Books, Open Library, Free E-Books.net etc. Basically, it does not own a digital library, but some e-resources can be accessed through some third-party site from this website.

## Concluding Remarks

Digital Library initiative was actually started in India at mid-1990s, but it got the pace at foremost after 2000s. When the Information Technology and Internet came prominently in Indian market, people actually started to think about the concept of digitization where in Europe there was already the concept of digitization and preservation the information as digitized version.

Although the digital library initiative was started in India in mid 1990s but very unexpected that till now there is not much development in the initiative for digitizing the information in the libraries of the North-Eastern region. Although, there are a couple of digital libraries in the NE region of India today, but it is observed in maximum place that users got trouble during the time of accessing the digital content. Perhaps, the main problem is the management of digital library due to the lack of knowledge of the particular library management. Also, many of those libraries only subscribed e-journals and provide users to get online access from any renowned website through the institution's website to get free e-content. Mainly digital libraries are seen in the academic institutions' websites are not active or working. So it can be pointed out that the lack knowledge and professionalism in India in the case of making a digital library.

There are many digital libraries (or, digital repository) in North-East India (mainly in the academic institutions), and they all claim that they have a digital library management system. However, there is a lack of proper library management system. Some have the different section for digital library in their particular website but there is no link in the section for accessing the digital library. Also, some have the digital library but the actual users' statistics from the website is not available.  Moreover, many libraries have the closed access for their digital collection.

Library networking by various agencies have been taking initiatives for digital library development in India. Amongst all other library Networks, INFLIBNET has been playing a significant role for the modernization of University and College Libraries of India in general and North-East Region of India in particular.

At present, India is not so advanced in terms of digital library but there has been some development. However, compared to the whole of India, there has been no significant development of digital libraries in north East India. So, there is needed a strong national policy that may focus not only in the digital data consistency in the library but also on the copyright issues and policies to obey the basic criteria about the information rights about digitaization to establish a digital library in India.

This study is based on the selected digital libraries in NE India. This is also based on the secondary information sources, particularly the websites on the respective websites. A comparative study of digital library initiatives of other parts of India will perhaps give a better and holistic picture of digital library initiatives in India.

**Notes: The following websites are access for this study during July to September 2022**

1. Rajiv Gandhi University Library available at: https://rgu.ac.in/library/ (accessed on 28th September 2022)
2. Virtual Library Solutions is the Reference Library of the Department of Cultural Affairs, Directorate of Research, Government of Arunachal Pradesh. Available at: http://referencelibraryresearch.com/historical-background/ (accessed on 28th September 2022
3. Arunachal University of studies https://www.arunachaluniversity.ac.in/pages/library (accessed on 18th September 2022)
4. Digital Library of Assam, Directorate of Library Services, Department of Cultural Affairs, Government of Assam https://publiclibraryservices.assam.gov.in/ (accessed on 18th September 2022)
5. Digital Library: Raj Bhavan, Assam http://rbassam.digitallibrary.co.in/ (accessed on 1th September 2022)
6. Assam District Library, Guwahati https://publiclibraryservices.assam.gov.in/ (accessed on 1th September 2022)
7. National Law University and Judicial Academy Digital Repository cum Digital Library http://librarynlujaa.blogspot.com (accessed on 20th August 2022)
8. Diphu Government College http://diphugovernmentcollege.com/DGC/CentralLibrary.action (accessed on 28th September 2022)
9. Central Institute of Technology, Kokrajhar http://centrallibrary.cit.ac.in/ (accessed on 20th August 2022)
10. Central Library, North-Eastern Hill University https://www.nehu.ac.in/lib rary/index.html (accessed on 1st August 2022)
11. Central Library: Nagaland University https://library.nagalanduniversity.ac.in/?q=Digital%20Libraries (accessed on 1st August 2022)
12. Teesta-Indus Central Library, Sikkim University https://library.cus.ac.in/index.php/subject-wise-e-books/ (accessed on 22nd July 2022)
13. Manipur University Library https://mulibrary.manipuruniv.ac.in/ (accessed on 22nd July 2022)
14. Manipur Technical University Library http://mtu.ac.in/library/ (accessed on 28th September 2022)
15. Mizoram Sate Library https://statelibrary.mizoram.gov.in/page/digital-library (accessed on 9th July 2022)
16. Birchandra State Central Library https://bcscl.tripura.gov.in/ (accessed on 9th July 2022)